\documentclass[12pt, a4paper]{article}

\usepackage{amsmath,amssymb}
\usepackage{times,fancyhdr}
\usepackage{graphicx}
\usepackage{amsfonts}

\begin{document}
\noindent Acta Biochim Biophys Sin 2013, 45(4): xxx-xxx, DOI: 10.1093/abbs/gmt031.\\
--------------------------------------------------------------------------------------------------\\

\noindent {\bf Review Article}\\

\vskip 0.3cm

\noindent {\bf {\Large Molecular Dynamics Studies on 3D Structures of the Hydrophobic Region PrP(109--136)}}\\

\vskip 0.3cm

\noindent {\bf {\small Jiapu Zhang$^{1*}$, Yuanli Zhang$^2$}}\\

\vskip 0.1cm

\noindent {\small
$^1$Graduate School of Sciences, Information Technology and Engineering, CIAO,
The University of Ballarat, MT Helen Campus, Victoria 3353, Australia;\\
$^2$School of Basic Medical Sciences,
Taishan Medical University, Shandong 271000, China.\\  
$^*$Correspondence address. Tel: +61-3-5327 6335; E-mail: j.zhang@ballarat.edu.au\\

\vskip 0.3cm

\noindent {\bf {\sl Abstract} Prion diseases, traditionally referred to as transmissible spongiform encephalopathies (TSE), are invariably fatal and highly infectious neurodegenerative diseases that affect a wide variety of mammalian species, manifesting as scrapie in sheep, bovine spongiform encephalopathy (BSE, or `mad-cow' disease) in cattle, and Creutzfeldt-Jakob disease (CJD), Gerstmann-Sträussler-Scheinker syndrome(GSS), Fatal familial insomnia (FFI) and Kulu in humans, etc. These neurodegenerative diseases are caused by the conversion from a soluble normal cellular prion protein (PrP$^{\text{C}}$) into insoluble abnormally folded infectious prions (PrP$^{\text{Sc}}$). The hydrophobic region PrP(109--136) controls the formation into diseased prions: the normal PrP(113--120) AGAAAAGA palindrome is an inhibitor/blocker of prion diseases ({\it Mol Cell Neurosci 15: 66--78}), and the highly conserved glycine-xxx-glycine motif PrP(119--131) can inhibit the formation of infectious prion proteins in cells ({\it J Biol Chem 285: 20213-–20223}). This article gives detailed reviews on the PrP(109--136) region and presents the studies of its three-dimensional structures and structural dynamics.}\\

\vskip 0.1cm

\noindent \textsl{Keywords:} hydrophobic region; PrP(109-136); AGAAAAGA palindrome; glycine-xxx-glycine motif; molecular dynamics study\\

\vskip 0.1cm

\noindent Received 08/01/2013; Accepted 18/02/2013;

\section{Introduction}
Prion diseases such as Creutzfeldt-Jakob disease (CJD) in humans and bovine spongiform encephalopathy (BSE or `mad-cow' disease) in cattle are invariably fatal neurodegenerative diseases. Prions differ from conventional infectious agents in being highly resistant to treatments that destroy the nucleic acids found in bacteria and viruses. The infectious prion is thought to be an abnormally folded isoform (PrP$^{\text{Sc}}$) of a host protein known as the prion protein (PrP$^{\text{C}}$). The conversion of PrP$^{\text{C}}$ to PrP$^{\text{Sc}}$ occurs post-translationally and involves conformational change from a predominantly $\alpha$-helical protein to one rich in $\beta$-sheet amyloid fibrils. Much remains to be understood about how the normal cellular isoform of the prion protein PrP$^{\text{C}}$ undergoes structural changes to become the disease associated amyloid fibril form PrP$^{\text{Sc}}$. The hydrophobic domain of PrP$^{\text{C}}$(109--136) is highly conserved, containing a palindrome and the repeats of the GxxxG protein-protein interaction motif (two glycines separated by any three residues; please note that the minimum number of residues to form fibrils should be 5 \cite{brown2000}). It is reported that the palindrome AGAAAAGA is an inhibitor/blocker of prion diseases \cite{brown2000, holscher_etal1998} and the glycine-xxx-glycine motif GAVVGGLGGYMLG is also an inhibitor of prion diseases \cite{harrison_etal2010, cheng_etal2011, lee_etal2008}. The alterations of residues in AGAAAAGA and GAVVGGLGGYMLG will drastically affect the ability of cells and lead to the amyloid fibril formations (e.g. A117V will cause the Gerstmann-Straussler-Scheinker prion disease, and the numerous mutants in \cite{harrison_etal2010}). Our computational results also confirm the amyloid fibril formation ability of the PrP(109--136) hydrophobic region (Figure 1).
\begin{figure}[h!] \label{Fig01_Energy-ResidueNumber}
\centerline{
\includegraphics[width=6.9in]{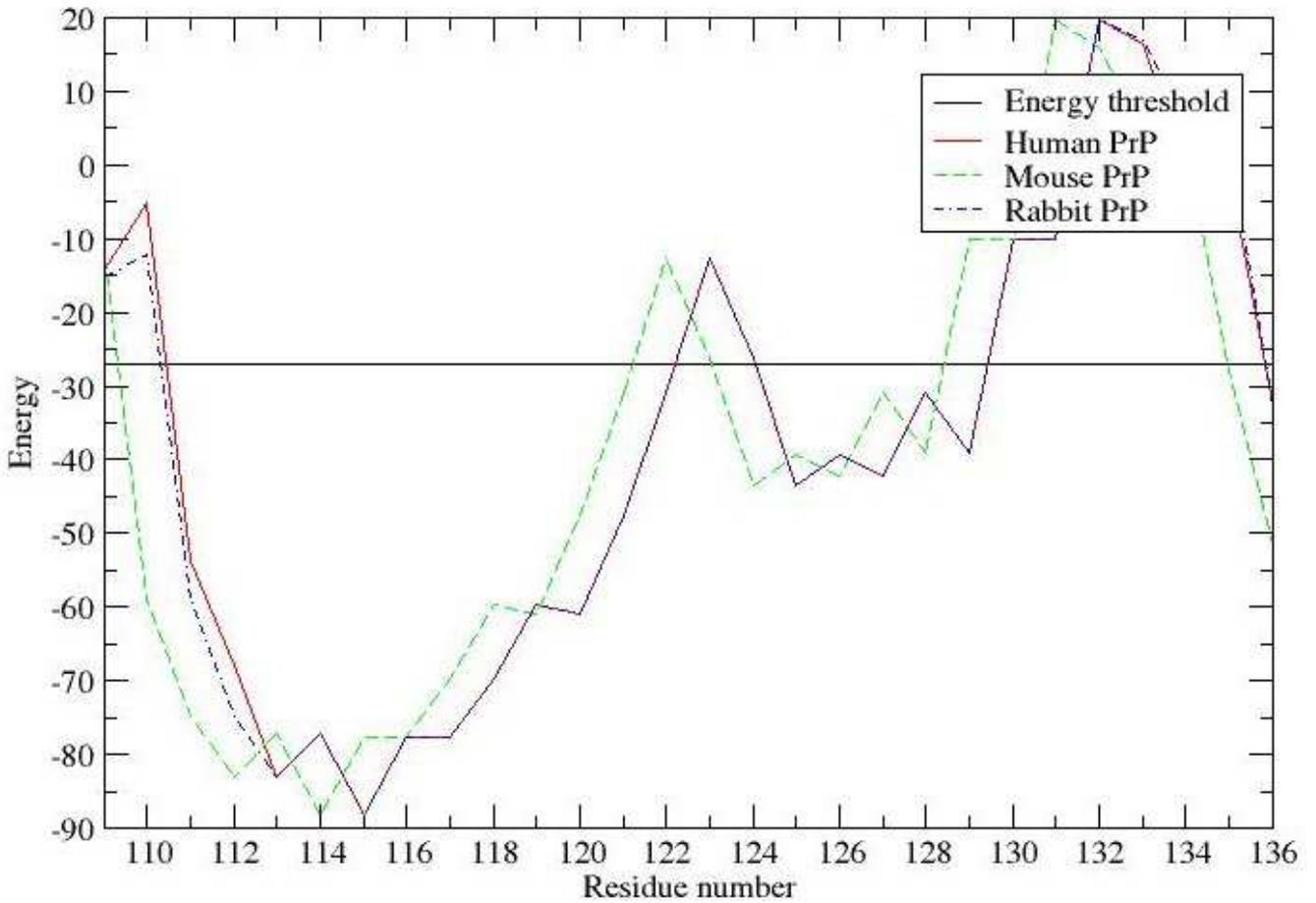}
}
\caption{{\it Identifying the amyloid fibril formation property in PrP(109--136) region by the fibril prediction program of \cite{zhang_etal2007}.}}
\end{figure} 
\noindent As shown in Figure 1, if energy is less than the threshold energy -26 kcal/mol then amyloid fibril is formed in the corresponding region of residues, thus, the palindrome segment PrP(113--120) and the GLGGY segment PrP(124--128) can be firmly confirmed having a strong amyloid fibril formation property. This paper will give detailed reviews on the PrP(109–-136) region from the 3D molecular structure (Figure 2) point of view and presents the studies of its molecular structural dynamics. 
\begin{figure}[h!] \label{Fig01_2LBG12_1QLX}
\centerline{
\includegraphics[width=6.9in]{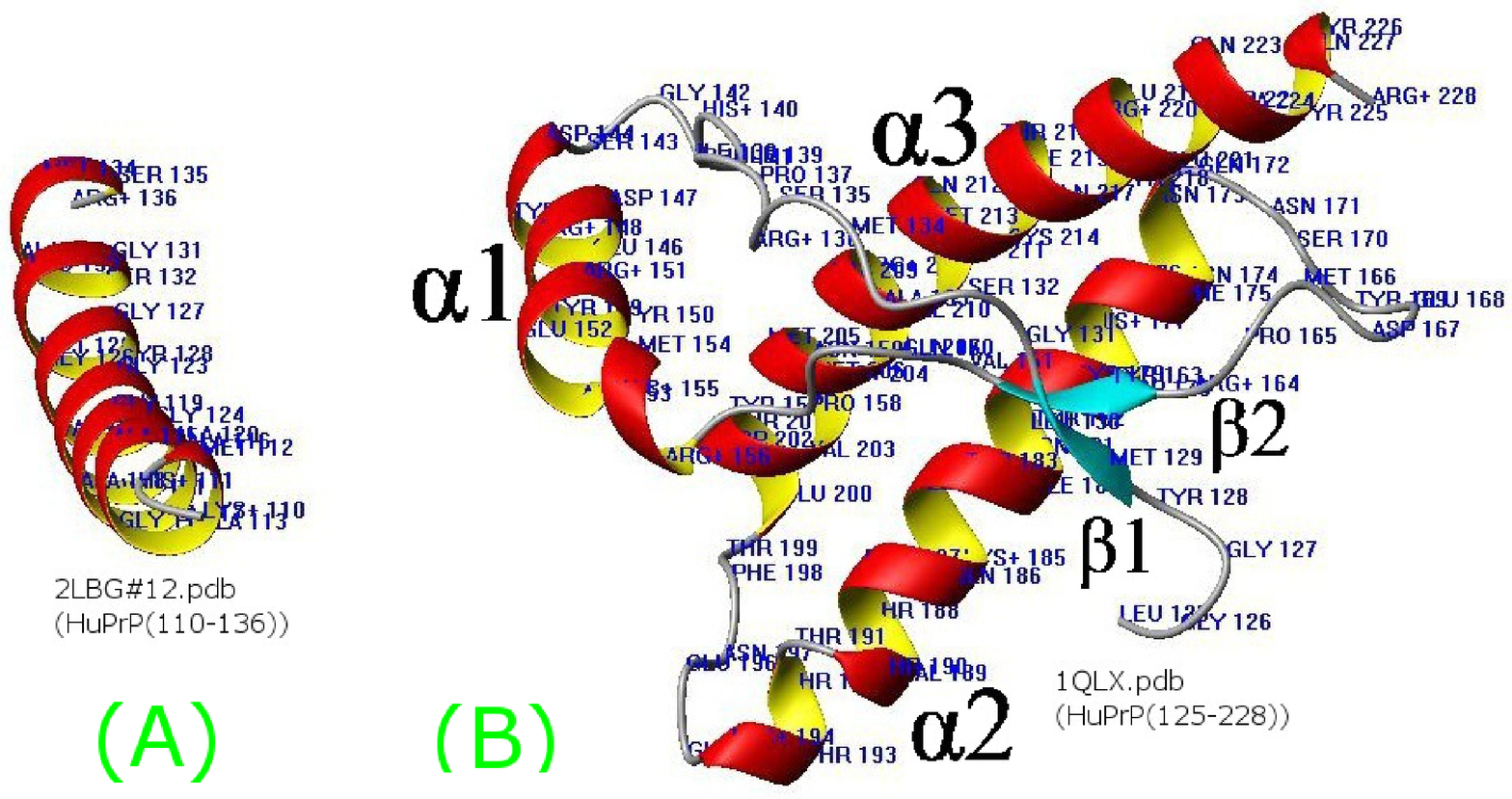}
}
\caption{{\it The 3D molecular structure of human PrP. (A): the 3D molecular NMR structure of human PrP(110–-136) in dodecylphosphocholine micelles. (B): the 3D molecular NMR structure of human PrP(125--228).}}
\end{figure} 
The rest of this paper is arranged as follows. Section 2 will give a survey of the research works on AGAAAAGA and present the 3D structure of amyloid fibrils in the AGAAAAGA segment. Then, in Section 3, the analysis of the glycine-XXX-glycine motif GAVVGGLGGYMLG inhibiting prion diseases will be done from the molecular structural point of view. Section 4 will make some concluding remarks on the hydrophobic PrP(109--136) region.

\section{Studies on the PrP(113--120) AGAAAAGA}
\subsection{A Survey of the Research Works on AGAAAAGA}
The highly conserved hydrophobic palindrome AGAAAAGA PrP(113--120) has been considered essential to PrP conformational conversion.  Firstly we give a survey of the research works on {\sl prion AGAAAAGA} listed in the PubMed database ({\sl www.ncbi.nlm.nih.gov}).

The C-terminally-truncated human prion protein variant Y145Stop (i.e. HuPrP(23-–144)) is a valuable model for understanding the fundamental properties of amyloid formation. To examine the role of AGAAAAGA segment in fibrillization of PrP(23-–144), a deletion variant $\Delta$(113-–120) PrP(23-–144) (in which the palindrome sequence is missing) is used \cite{jones_etal2011}. The deletion results in an altered amyloid $\beta$-core without affecting amyloidogenicity or seeding specificity; this concludes that the core of some amyloids contains ``essential" (nucleation-determining) and ``nonessential" regions, with the latter being flexible in amino acid sequence space \cite{jones_etal2011}.

The amyloid fibrils formed by the polypeptides of PrP(113--127), AGAAAAGAVVGGLGG, are taken as the model compound to investigate the biophysical principles governing the steric zipper amyloid fibril formation \cite{cheng_etal2011, lee_etal2008, sawaya_etal2007, kajava_etal2006}. The target fibrils adopt the structural motif of Class 7 steric zipper, which is formed by stacking of antiparallel $\beta$-sheet layers with residue $117+k$ forming backbone hydrogen bonds to residue $120-k$ \cite{cheng_etal2011, lee_etal2008, sawaya_etal2007}.

Computer simulations of amyloid fibril formation by the Syrian hamster prion protein (SHaPrP) residues AGAAAAGA, the mouse prion protein (MoPrP) residues VAGAAAAGAV, and their variations GA$_6$G (a longer uninterrupted Alanine stretch flanked by Glycine), (AG)$_4$ (a complete disruption of hydrophobic residues), A$_8$, GAAAGAAA (a mimic of A$\beta$(29-–36)), A$_{10}$, V$_{10}$, GAVAAAAVAG (uninterrupted hydrophobic sequence), VAVAAAAVAV (less flexible than MoPrP(111-–120)) are studied in \cite{wagoner_etal2011}. The first two peptides are thought to act as the velcro that holds the parent prion proteins together in amyloid structures and can form fibrils themselves \cite{wagoner_etal2011}.

AGAAAAGA of HuPrP(105--210) is reported being involved into the oligomerization and engaged in intra- and/or inter-molecular interactions \cite{sasaki_etal2008}. The SHaPrP(109--122) peptide near the AGAAAAGA region is observed to form steric zipper fibrils by the data of crystal solid-state NMR and molecular dynamics (MD) \cite{lee_etal2008}.

The cellular isoform of the prion protein PrP$^{\text{C}}$ is located at the cell membrane. Studies have shown that exposure of cells to copper (Cu) causes internalisation of PrP$^{\text{C}}$ in vitro, and deletion mutation studies have shown that the palindromic region, amino acids 113-–120 with the sequence AGAAAAGA is essential for copper-induced internalisation to occur \cite{haigh_etal2005}. Kourie et al (2003) studied the copper modulation of ion channels of PrP(106--126) mutant prion peptide fragments and found that the hydrophobic core AGAAAAGA is not a Cu$^{\text{2+}}$-binding site (but at Met 109 and His ll1 ion channels can be formed) \cite{kourie_etal2003}.

The AGAAAAGA palindrome in PrP is required not only for the attainment of the PrP$^{\text{Sc}}$ conformation but also to generate a productive PrP$^{\text{Sc}}$-PrP$^{\text{C}}$ complex that leads to the propagation of PrP$^{\text{Sc}}$ \cite{norstrom-and-mastrianni2005}. In contrast to wild type (wt) PrP, PrP lacking the palindrome (PrP $\Delta$(112–-119)) neither converted to PrP$^{\text{Sc}}$ nor generated proteinase K-resistant PrP. In \cite{norstrom-and-mastrianni2005}, we also know that synthetic peptides corresponding to the so-called ``toxic peptide" PrP(106-–126) segment form fibrils in solution with $\beta$-sheet structure, suggesting that PrP(106--126) segment may feature in PrP$^{\text{Sc}}$-PrP$^{\text{C}}$ associations, and larger peptides of PrP(107-–142) antagonize the in vitro conversion of PrP$^{\text{C}}$ to the protease-resistant state in a cell-free conversion model.

Zanuy et al (2003) reported that, for AGAAAAGA, the antiparallel strand orientation is preferred within the sheets but the parallel orientation is preferred between sheets \cite{zanuy_etal2003, ma_etal2002}. AGAAAAGA is one of the most highly amyloidogenic peptides and oligomers of AGAAAAGA were found to be stable when the size is 6 to 8 (hexamer to octamer) \cite{ma_etal2002}. Here the AGAAAAGA model of \cite{ma_etal2002} used for their MD is a homology structure.

In \cite{wegner_etal2002}, the following bioinformatics on AGAAAAGA is known. AGAAAAGA displays the highest tendency to form amyloid. Peptides containing AGAAAAGA are toxic to neurons in culture, whereby the sequence AGAAAAGA was found to be necessary but not sufficient for the neurotoxic effect \cite{brown2000}. The synthetic peptides derived from the central part of PrP$^{\text{C}}$(106--141) have an inhibitory effect on this conversion of PrP$^{\text{C}}$ to PK-resistant PrP$^{\text{Sc}}$. The presence of residues 119 and 120 (the two last residues within the motif AGAAAAGA) seems to be crucial for this inhibitory effect. Mutant PrP molecules carrying deletions of amino acids 108--121 or 114--121 are not convertible to PrP$^{\text{Sc}}$; therefore, the central hydrophobic region, spanning all or most of the sequence AGAAAAGA, plays an important role in the PrP$^{\text{Sc}}$-PrP$^{\text{C}}$ conversion process. Wegner et al (2002) assessed the effect of mutations at and around the AGAAAAGA hydrophobic sequence on protease resistance and found that mutations in the central AGAAAAGA hydrophobic region lead to immediate alterations in PrP structure and processing \cite{wegner_etal2002}.

The amyloidogenic and hydrophobic core AGAAAAGA has been implicated in modulation of neurotoxicity and the secondary structure of PrP(106–-126) \cite{kourie2001, jobling_etal1999}, which is dependent on the formation of aggregated fibril structures regulated by the AGAAAAGA core \cite{kourie2001}. 

Brown (2000) reported that AGAAAAGA blocks the toxicity of PrP(106-–126), suggesting that this sequence is necessary (but insufficient) for the interaction of PrP(106-–126) with neurons and targeting or use of the AGAAAAGA peptide may represent a therapeutic opportunity for controlling prion disease \cite{brown2000}.

HuPrP(106-–126) has been shown to be highly fibrillogenic and toxic to neurons in vitro. Jobling et al (1999) found that the AGAAAAGA PrP(113--120) hydrophobic core sequence is important for PrP(106-–126) toxicity probably by influencing its assembly into a neurotoxic structure and the hydrophobic sequence may similarly affect aggregation and toxicity observed in prion diseases \cite{jobling_etal1999}.

Chabry et al (1998) reported that peptides from the central part of the hamster PrP 106--141 (where residues in the vicinity of positions 106–-141 of PrP$^{\text{Sc}}$ and/or PrP$^{\text{C}}$ are critically involved in the intermolecular interactions that lead to PrP$^{\text{Sc}}$ formation) could completely inhibit the conversion induced by preformed PrP$^{\text{Sc}}$ and the presence of residues 119 and 120 from the highly hydrophobic sequence AGAAAAGA (PrP(113--120)) was crucial for an efficient inhibitory effect \cite{chabry_etal1998}.

Holscher et al (1998) reported that the presence of AGAAAAGA of mouse PrP$^{\text{C}}$ plays an important role in the conversion process of PrP$^{\text{C}}$ into PrP$^{\text{Sc}}$ and that a deletion mutant lacking these codons indeed behaves as a dominant-negative mutant with respect to PrP$^{\text{Sc}}$ accumulation \cite{holscher_etal1998}.

PrP AGAAAAGA is the most highly amyloidogenic peptide and is conserved across all species \cite{gasset_etal1992, govaerts_etal2004}. Gasset et al (1992) reported there are similarities between the PrP sequence AGAAAAGA and that of silkworm fibroin, and the homology between PrP sequence AGAAAAGAVVGGLGG and that of spider fibroin \cite{gasset_etal1992}. Thus, we may say that the hydrophobic region of PrP(109--136) should be a region to form $\beta$-sheets and amyloid polymers, instead of $\alpha$-helices of the Garnier-Robson analysis \cite{garnier_etal1978}.

\subsection{3D Structure of Prion AGAAAAGA Amyloid Fibrils}
Seeing the above survey, we know that prion AGAAAAGA peptide has been reported to own an amyloid fibril property (initially described in 1992 by Gasset et al of Prusiner's Group). However, there has not been traditional X-ray or NMR experimental structural bioinformatics for this octapeptide yet, due to the unstable, noncrystalline and insoluble nature of this region, which just falls within the N-terminal unstructured region of prion proteins. Studies on atomic-resolution structures of the AGAAAAGA peptide will prove useful in future experimental studies on this region, aspects of the structure or the dynamics of this region should play a role in the aggregation process, and knowledge of these may be useful for the goals of medicinal chemistry for controlling prion diseases. Zhang (2011) successfully constructed three amyloid fibril models (denoted as Models 1--3) for the PrP(113--120) AGAAAAGA region \cite{zhang2011}. Using all the PDB templates of \cite{sawaya_etal2007} (with PDB IDs: 2OKZ, 2ONW, 2OLX, 2OMQ, 2ON9, 2ONV, 2ONA, 1YJO, 2OL9, 2OMM, 2ONX, 2OMP, 1YJP), simulation conditions are completely consistent with the experimental works of \cite{sawaya_etal2007} and the simulation methods of package Amber 10 \cite{amber10} are: the optimization steepest descent (SD) method and conjugate gradient (CG) method to relax the Models, then the standard simulated annealing method to make the equilibration of Models sufficiently stable and then the SD-CG optimization methods to refine the Models.      
\begin{figure}[h!] \label{Fig02_model1}
\centerline{
\includegraphics[width=6.9in]{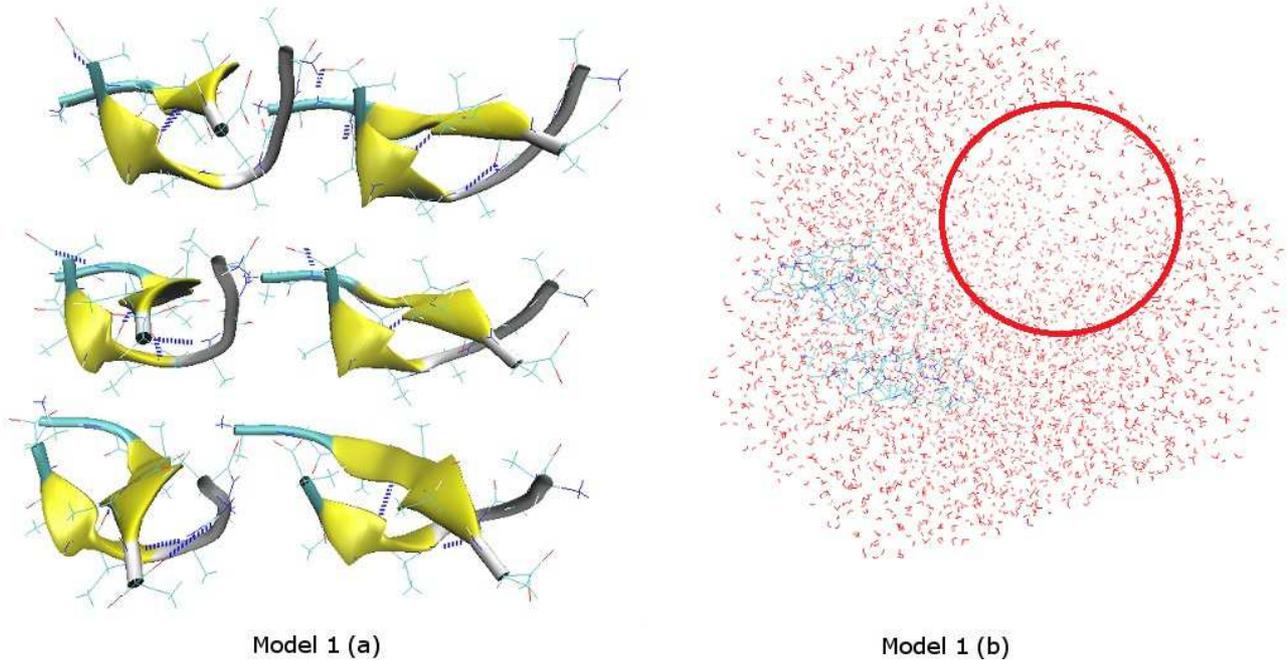}
}
\caption{{\it Model 1. (A): blue dashed lines denote the hydrogen bonds between the pairs of $\beta$-strands, and the $\beta$-sheets are maintained by van der Waals contacts and hydrophobic packings. (B): the red circle denotes there are very few water molecules in this part of the truncated octahedral box of TIP3P waters, the trajectories of the movement of the amyloid fibril formulate a ``cage" made of waters and the amyloid fibril flies to the bottom of the ``cage", the mouth of the ``cage' is open -- this illuminates to us the amyloid fibril is very hydrophobic.}}
\end{figure}
\begin{figure}[h!] \label{Fig03_model2}
\centerline{
\includegraphics[width=6.9in]{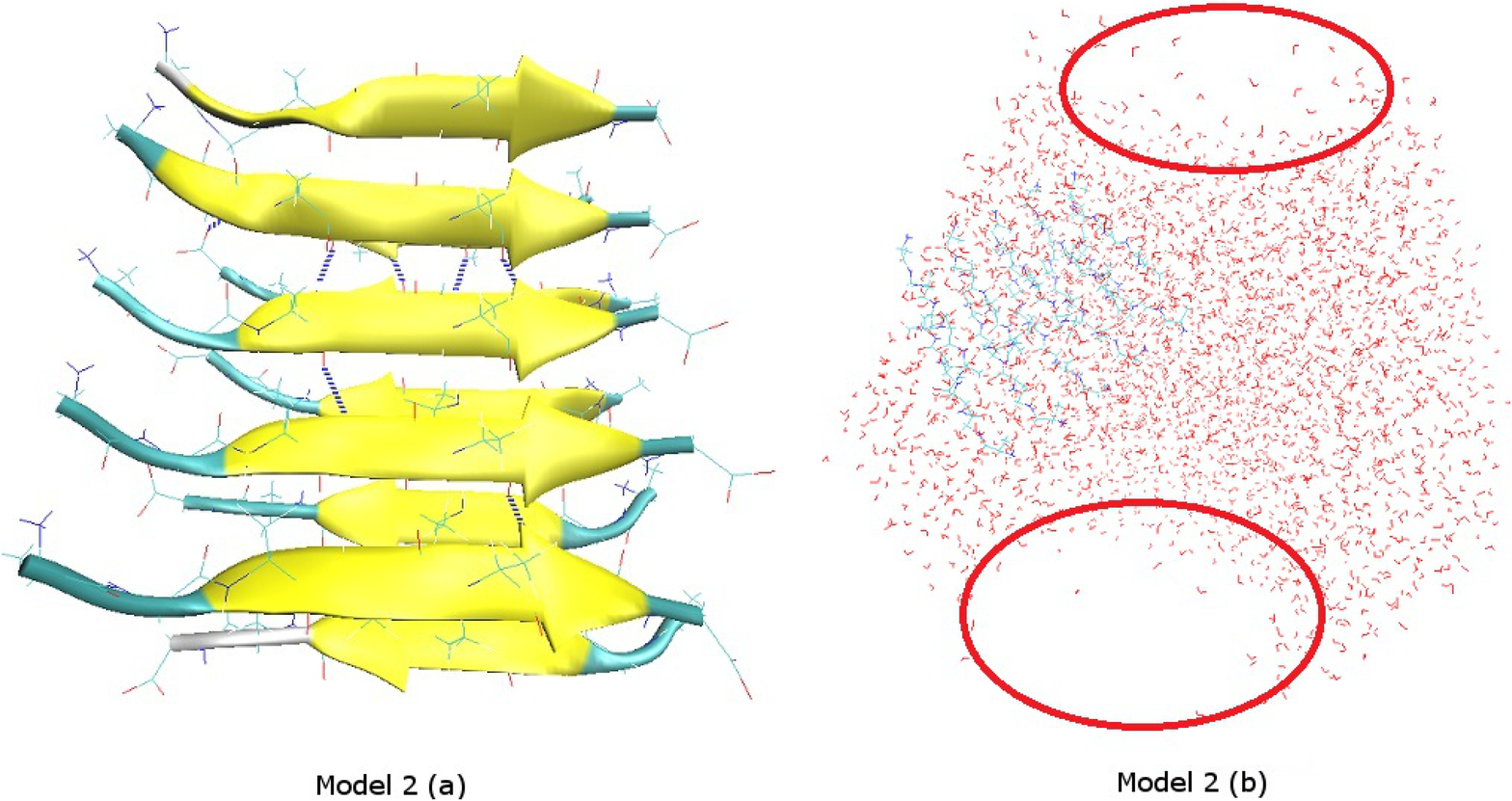}
}
\caption{{\it Model 2. (A): blue dashed lines denote the hydrogen bonds between the pairs of $\beta$-strands, and the $\beta$-sheets are maintained by van der Waals contacts and hydrophobic packings. (B): the red half-circles denote there are very few water molecules in this parts of the truncated octahedral box of TIP3P waters, the trajectories of the movement of the amyloid fibril formulate a ``cage" made of waters with two open ``mouths" circled in red -- this illuminates to us that the hydrophobic property of amyloid fibrils can form a very interesting movement pattern of trajectories.}}
\end{figure}
\begin{figure}[h!] \label{Fig04_model3}
\centerline{
\includegraphics[width=6.9in]{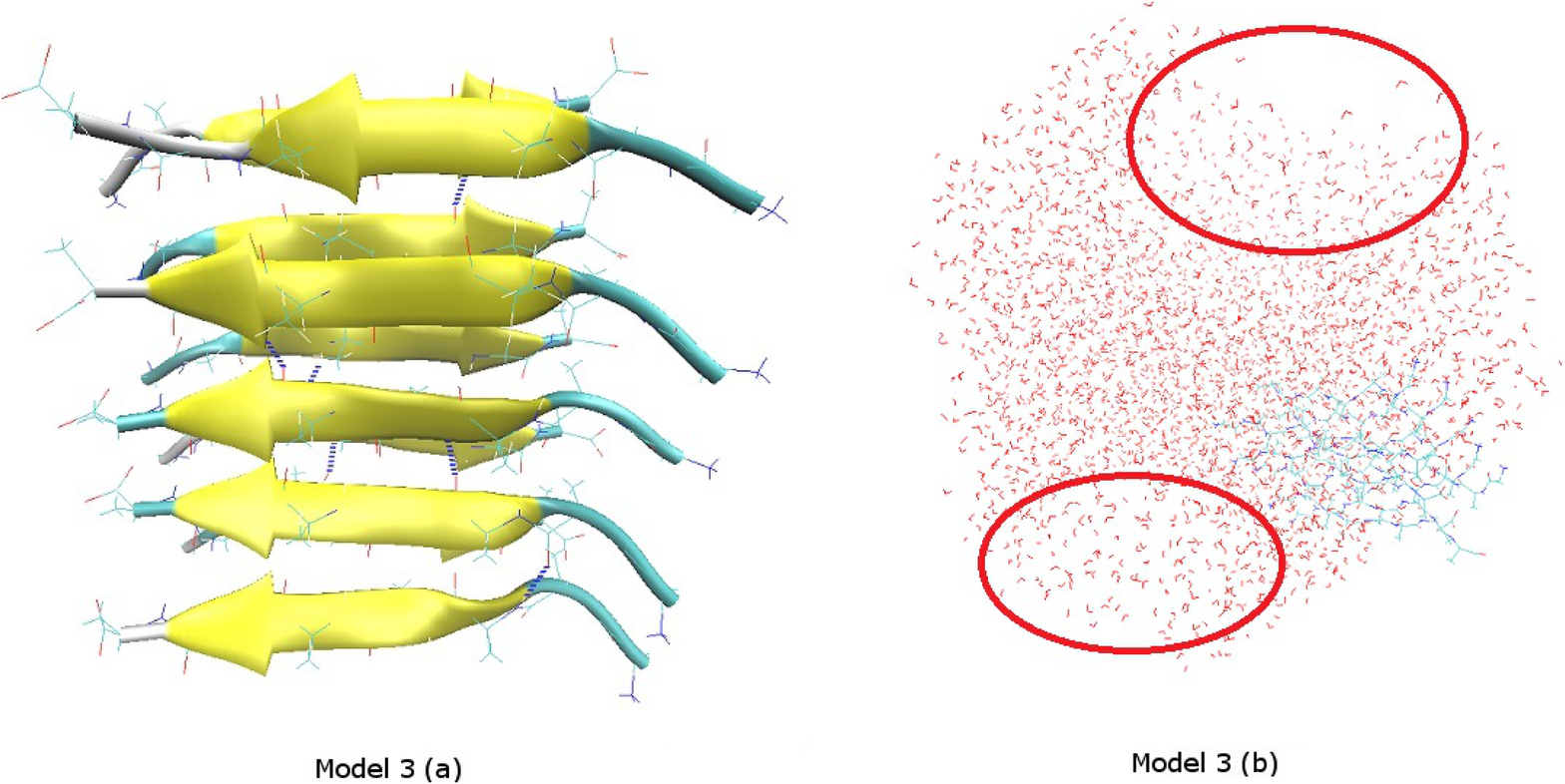}
}
\caption{{\it Model 3. (A): blue dashed lines denote the hydrogen bonds between the pairs of $\beta$-strands, and the $\beta$-sheets are maintained by van der Waals contacts and hydrophobic packings. (B): the red half-circles denote there are very few water molecules in this parts of the truncated octahedral box of TIP3P waters, the trajectories of the movement of the amyloid fibril formulate a ``cage" made of waters with two open ``mouths" circled in red -- this illuminates to us that the hydrophobic property of amyloid fibrils can form a very interesting movement pattern of trajectories, and this also illuminates us the property of periodic boundary calculations in PMEMD (particle mesh eward molecular dynamics) simulations.}}
\end{figure}
\noindent Model 1 and Models 2--3 belong to Class 7 and 1 of \cite{sawaya_etal2007} respectively, i.e. Model 1 is $\beta$-strand antiparallel, face=back, up-up (Figure 3) (where numerical results show the agreement with \cite{cheng_etal2011, ma_etal2002}), and Models 2--3 are $\beta$-strand parallel, face-to-face, up-up (Figures 4--5). In all these models, there is about 5 angstroms between the two closest adjacent $\beta$-sheets, maintained by hydrophobic bonds, and about 4.5 angstroms between the two closest adjacent $\beta$-strands, which are linked by hydrogen bonds. Illuminated by PDB templates 3FVA, 3NHC/D, 3NVF/G/H/E, 3MD4/5, 2OMP, computational approaches of global optimization, local search energy minimization (EM), simulated annealing (SA) and structural bioinformatics etc or introducing novel mathematical formulations and physical concepts into molecular biology may allow us to obtain a description of the protein 3D structure at a submicroscopic level for prion AGAAAAGA amyloid fibrils \cite{zhang2011, zhang2011a, zhang_etal2011a, zhang_etal2011b, zhang2012, zhang_etal2012a}. 

\section{Structural Studies on the PrP(119--131) GAVVGGLGGYMLG}
Some 30 segments from the Alzheimer's amyloid-$\beta$ (A$\beta$) and tau proteins, the PrP prion protein, insulin, etc form amyloid-like fibrils, microcrystals that reveal steric zipper structures \cite{sawaya_etal2007}. Harrison et al (2006) reported there are similarities between A$\beta$ and PrP in the segment of the three GxxxG repeats (where both A$\beta$ and PrP have the crucial residue Methionine located in the middle (GxMxG) of the last repeat) \cite{barnham_etal2006, harrison_etal2007} that controls prion formation \cite{harrison_etal2010}. Harrison et al (2010) used cell biological approaches of investigating numerous mutants in this region to reveal the mechanism of prion inhibition, and mutagenesis studies demonstrate that minor alterations to this highly conserved region of PrP$^{\text{C}}$ drastically affect the ability of cells to uptake and replicate prion infection in both cell and animal bioassay \cite{harrison_etal2010}. This section presents some explanations for the biological experimental performance of \cite{harrison_etal2010} from the molecular structural point of view.

First, we do the alignments of the structured region of mouse, human, dog, rabbit, horse and elk PrP$^{\text{C}}$ (with PDB IDs 1AG2, 1QLX, 1XYK, 2FJ3, 2KU4, 1XYW respectively) (Figure 6). Figure 6(A) shows the PrP(125--136) 3D structures of human, dog, rabbit, horse and elk superposed onto mouse with backbone-atom-RMSD values 2.579427, 2.228940, 2.745877, 2.532690, 2.877734 angstroms respectively. In Figure 6(B), the alignment of sequences was generated by the online CLUSTAL 2.1 program at {\sl http://www.ebi.ac.uk/Tools/msa/clustalw2/}, where ``*" means that the residues or nucleotides in that column are identical in all sequences in the alignment, ``:" means that conserved substitutions have been observed, ``." means that semi-conserved substitutions are observed, the red colored residues are Small (small+ hydrophobic (incl.aromatic -Y)), the blue colored residues are Acidic, the green colored residues are Hydroxyl + sulfhydryl + amine + G, and the gray colored residues are Unusual amino/imino acids etc; the residue numbers are from 119 to 231 (for rabbit the numbers are 1 less than others); the last column of numbers denotes the number of residues accounted from 1; the PrP 125--136 LGGYMLGSAMSR, $\beta$-strand 1 ({\bf $\beta$1}), $\alpha$-helix 1 ({\bf $\alpha$1}), $\beta$-strand 2 ({\bf $\beta$2}), $\alpha$-helix 2 ({\bf $\alpha$2}) and $\alpha$-helix 3 ({\bf $\alpha$3}) of a PrP structure (Figure 2(B)) were underline denoted. 
%\begin{figure*}[h!] 
%\centerline{
%\includegraphics[width=4.0in]{Fig05_alignments1.eps}
%}
%\end{figure*} 
\begin{figure}[h!] \label{Fig05_alignments}
\centerline{
\includegraphics[width=6.2in]{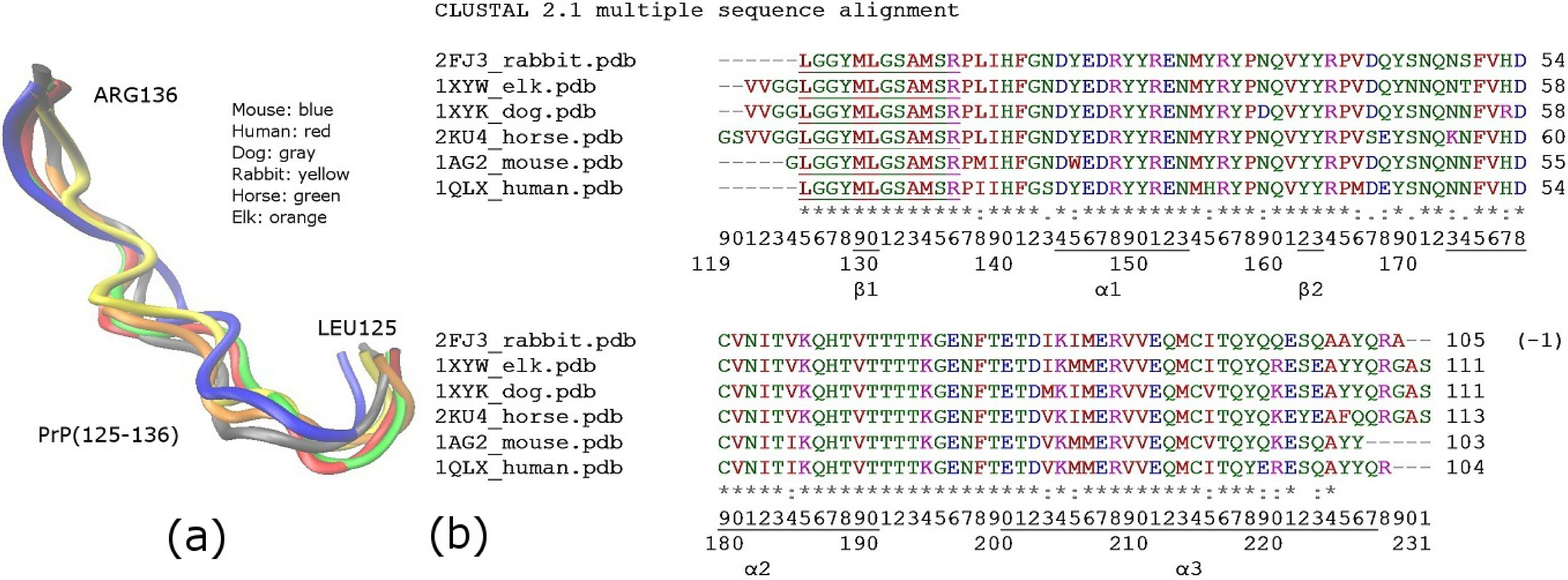}
}
\caption{{\it The alignments of the structured region of mouse, human, dog, rabbit, horse and elk proteins with PDB IDs 1AG2, 1QLX, 1XYK, 2FJ3, 2KU4, 1XYW respectively. (A) shows the 3D structural alignments of PrP$^{\text{C}}$(125--136) (whose residues were underlined in (B)). (B) shows the 1D sequence alignments of PrP$^{\text{C}}$(119--231).}}
\end{figure} 
In Figure 6, we can see the PrP(125--136) LGGYMLGSAMSR is highly conserved among all species, strongly suggesting it has functional and evolutionary significance. To understand the functions of proteins at a molecular level, in protein structures, the non-covalent interactions such as hydrogen bonding (HB), ionic interactions (SB), van der Waals forces (vdW), and hydrophobic packing (HP) are driving the proteins to be able to perform their biological functions ({\sl http://en.wikipedia.org/wiki/Protein\_structure}). Thus, in the below, we will investigate the HBs, SBs, vdWs, HPs in the structures and their structural dynamics of  
      LGGYMLGSAMSR of human and rabbit,
     GLGGYMLGSAMSR of mouse,
  VVGGLGGYMLGSAMSR of elk and dog,
GSVVGGLGGYMLGSAMSR of horse (where S120 is special for horse, instead of A120)
and structural connections with other residues/loops/sheets/helices in the C-terminal. Mutations will destroy these non-covalent interactions that well maintain the structure so the function of the prion protein. This will give clear explanations to the mutants in the Glycine-rich region of \cite{harrison_etal2010, choi_etal2006}, which affect the uptake of prion infectivity very much. 

In 2010, horses were reported to be resistant to prion diseases \cite{khan_etal2010}. First, we analyze the role of GSVVGGLGGYMLGSAMSR (PrP(119--136)) in horse PrP$^{\text{C}}$(119--231) (PDB entry 2KU4). In our study \cite{zhang2011b}, the MD simulation conditions are at 350 K in explicit solvent under neutral and low pH environments, heatings are using the Langevin thermostat algorithm in constant NVT ensembles and the equilibrations and productions are using Langevin thermostat algorithm in constant NPT ensembles. Seeing the Tables I--III of the Supplementary Material of \cite{zhang2011b}, we know the following HBs of GSVVGGLGGYMLGSAMSR (PrP(119--136)):
\begin{enumerate} 
{\small
\item[] ARG136-TYR157 (linking $\beta$1-to-$\alpha$1 loop with $\alpha$1-to-$\beta$2 loop, 68.27\%, 71.04\%),
\item[] MET134-ASN159 (linking $\beta$1-to-$\alpha$1 loop with $\alpha$1-to-$\beta$2 loop, 29.83\%, 21.47\%),
\item[] GLY131-GLN160 (linking $\beta$1 with $\beta$2, 26.67\%, 37.03\%), 
\item[] SER132-GLN217 (linking $\beta$1 with $\alpha$3, 29.70\%, 12.90\%),
\item[] ARG136-PRO158 (linking $\beta$1-to-$\alpha$1 loop with $\alpha$1-to-$\beta$2 loop, 9.53\%, 6.01\%),
\item[] ARG136-TYR157 (linking $\beta$1-to-$\alpha$1 loop with $\alpha$1-to-$\beta$2 loop, 68.27\%, 71.04\% ),
\item[] GLY126-ARG164 (linking bend before $\beta$1 with $\beta$2-to-$\alpha$2 loop, 33.31\%$^1$),
\item[] TYR128-ASP178 (linking coil before $\beta$1 with $\alpha$2,  11.81\%$^1$),
\item[] SER120-LEU125 (in the peptide, 9.72\%$^1$), 
\item[] LEU125-10TYE128 (in the peptide, 5.58\%$^2$),
\item[] GLY119-4VAL122 (in the peptide, 6.51\%$^2$),
\item[] GLY127-ARG164 (link bend before $\beta$1 with $\beta$2-to-$\alpha$2 loop, 5.87\%$^2$),
}
\end{enumerate}
where the first percentage is for seed1 (\%$^1$) and the second percentage is for seed2 (\%$^2$) and the two seeds mean two different initial velocities of 30 ns MD, and the following HPs of GSVVGGLGGYMLGSAMSR (PrP(119--136)):
\begin{enumerate} 
{\small
%\item[] TYR162-LEU130 (100\%, 99.06\%) which is just linking $\beta$1 and $\beta$2 (we noticed that there is no HB between TYR162 and LEU130),
\item[] In PrP(119--136) of 2LBG.pdb, between the two adjacent residues there are always occupied by HPs with rate of 100\%, except for between GLY123 and GLY124, and between GLY126 and GLY127, 
\item[] Among the residues in PrP(119--136), there are HPs in
\begin{enumerate}
\item[] TYR128--LEU130 (where LEU130 is a residue in $\beta$1), 
\item[] VAL121--GLY119, GLY123, TYR128, MET129, LEU130 (where MET129 and LEU130 are in $\beta$1), 
\item[] VAL122--SER120, GLY124, LEU125, TYR128,
\item[] GLY123--SER120, LEU125,                       
\item[] TYR128--GLY126,                         
\item[] MET129--GLY131  (where both MET129 and GLY131 are in $\beta$1),                        
\item[] MET134--SER132, ARG136 (where SER132 is in $\beta$1),                       
\end{enumerate}
\item[] The HPs between a residue in PrP(119--136) and a residue in PrP(137--231):
\begin{enumerate}
\item[] SER135--PRO137 (where SER135, PRO137 is in $\beta$1-to-$\alpha$1 loop),
\item[] ARG136--MET154, TYR157 (where ARG136 is in $\beta$1-to-$\alpha$1 loop, MET154 and TYR157 are in $\alpha$1-to-$\beta$2 loop),
\item[] ASN159--SER135, ALA133, ARG136, MET134 (where ASN159 is in $\alpha$1-to-$\beta$2 loop), 
\item[] TYR162--TYR128, MET129, LEU130 (where MET129, LEU130 are in $\beta$1, and TYR162 is in $\beta$2, there is no HB between TYR162 and LEU130),
\item[] GLN217--GLY131 (where is in $\beta$1, GLN217 is in $\alpha$3), 
\item[] ARG164--GLY127, MET129, TYR128 (where MET129 is in $\beta$1, ARG164 is in $\beta$2-to-$\alpha$2 loop),
\item[] TYR163--TYR128, LEU130, MET129 (where MET129, LEU130 are in $\beta$1, TYR163 is in $\beta$2),
\item[] GLN162--ALA133, LEU130, MET134, GLY131, SER132 (where ALA133, LEU130, GLY131, SER132 are in $\beta$1, GLN162 is in $\beta$2),
\item[] VAL161--GLY131, LEU130 (where GLY131, LEU130 are in $\beta$1, VAL161 is in $\beta$2),
\item[] MET213--MET134 (MET213 is in $\alpha$3, MET134 is in the $\beta$1-to-$\alpha$1 loop),
\end{enumerate}
}
\end{enumerate}
with $>$ 50\% occupancy rate over the long MD trajectory of 30 ns. The mutations listed in \cite{harrison_etal2010} will break these HBs and HPs to lose the inhibition to prion diseases of horses.

Rabbits are also resistant to infection from prion diseases from some species \cite{vorberg_etal2003, lin_etal2011, li_etal2007, wen_etal2010a, wen_etal2010b} and the outbreak of ``mad rabbit disease" is unlikely \cite{fernandez-borges_etal2012, chianini_etal2012}. Nisbet et al (2010) reported there is 87\% sequence homology between mouse PrP and rabbit PrP, approximately 9 of the 33 (i.e. 1/3) of the difference in the region DGRRSSSTV of mouse and QRAAGVL of rabbit, and  residues surrounding the glycosylphosphatidylinositol anchor attachment site of PrP modulate prion infection \cite{nisbet_etal2010}. Under the MD simulation conditions of 450 K in explicit solvent, neutral and low pH environments, with heatings using the Langevin thermostat algorithm in constant NVT ensembles and the equilibrations and productions using Langevin thermostat algorithm in constant NPT ensembles, we have found rabbit PrP has the C-terminal residue R227 forming a HB/SB network with inner residues and the beginning residues of rabbit homology strcture PrP(120--229) (6EPA.pdb) and NMR structure PrP(124--228) (2FJ3.pdb) but mouse PrP has no such an Arginine residue at the end of C-terminal owning this property \cite{zhang2011c, zhang2011d}.
                             
Dogs were reported in 2008 to be resistant to infection from prion diseases from other species \cite{polymenidou_etal2008}. In our recent work \cite{zhang_etal2012b}, MD studies were done under 450 K in explicit solvent, neutral and low pH environments, with heatings using the Langevin thermostat algorithm in constant NVT ensembles and the equilibrations and productions using Langevin thermostat algorithm in constant NPT ensembles. In the hydrophobic region 121--136 of dog PrP, MD studies find that there are strong HBs S132--Y163--Y128 linking the two antiparallel $\beta$-strands, and strong HPs M134--A133, M129--L130, V121--V122 residing in the core of the hydrophobic region 121-136 \cite{zhang_etal2012b}. We should note that residue 129 for dog PrP is L129 (http://www.uniprot.org/uniprot/O46501) but for many others is M129.\\
                    
For the hydrophobic region PrP(109--136), we should also notice the following points: 
\begin{enumerate} {\small
\item[] (1) the segment GYMLGS or GYVLGS of HuPrP(127--132) can form Class 8 antiparallel amyloid fibrils (3NHC.pdb and 3NHD.pdb) \cite{sawaya_etal2007, apostol_etal2010}, 
\item[] (2) the mutants A117V and M129V cause GSS \cite{mastrianni_etal1995, daidone_etal2011, chen_etal2010} and CJD \cite{klemm_etal2012, wadsworth_etal2004} respectively,
\item[] (3) the $\beta$-sheet core of PrP$^{\text{Sc}}$ consists of three layers of $\beta$-strands E1(116--119), E2(129--132) and E3(160--164) \cite{demarco_etal2004}, where E1 and E2 are in the hydrophobic region (109--136),  
\item[] (4) H111 is a residue of copper binding sites \cite{rivillas-Acevedo_etal2011, jones_etal2005, klewpatinond_etal2007, hasnain_etal2001}, 
\item[] (5) Y128 is in the center of HB-and-SB-network of HB(Y128--D178), SB(D178--R164), HB(Y128--R164), HB(Y128--H177), HB(H177-N154)  \cite{barducci_etal2006, barducci_etal2005, alonso_etal2001, glockshuber_etal1998},
\item[] (6) A133 and S132 have HBs with R220 and a water binding site with G131 \cite{knolmurodov_etal2003, desimone_etal2005}, in PrP(113--132) the hydrophobic cluster with vdWs rendering of atoms in residues 113--127 interacts with the first $\beta$-strand (see Figure 2(B) of \cite{james_etal1997}), M129 makes interactions with the side chain of V122 and pulls the N-terminus into the $\beta$-sheet \cite{alonso_etal2001} and M129 is very close to Y163 \cite{barnham_etal2006},
\item[] (7) conservation of the Gly-rich region PrP(119--131) is required for uptake of prion infectivity \cite{harrison_etal2010, peretz_etal1997, yuan_etal2005} is shown by the physical or chemical properties of numerous mutations in the PrP(119--131) Gly-rich region \cite{harrison_etal2010}, 
\item[] (8) one O-linked sugar at Ser135 can affect the coil-to-$\beta$ structural transition of the prion  peptide, but at Ser132 the effect is opposite \cite{chen_etal2002}, etc., and
\item[] (9) the NMR structure of HuPrP(110--136) in dodecylphosphocholine micelles was known (2LBG.pdb) \cite{sauve_etal2012} and we do MD simulations for it as follows.
}
\end{enumerate}

We used the ff03 force field of the AMBER 11 package \cite{amber11}, in a neutral pH environment. The systems were surrounded with a 12 angstroms' layer of TIP3PBOX water molecules and with 2 Cl- ions added using the XLEaP module of AMBER 11. The templates used are the 2LBG.pdb from the Protein Data Bank and its 17 mutants at G114V, A117V, G119A, G119L, G119P, A120P, G123A, G123P, G124A, L125A, G126A, G127A, G127L, M129V, G131A, G131L, G131P, which are got by the mutate module of the free package Swiss-PdbViewer Version 4.1.0 (http://spdbv.vital-it.ch/). These 18 models were firstly optimized by SD method and then CG method for 2 stages. Minimization Stage 1 is holding the solute fixed with a force constant of 500.0 kcal mol$^{\text{-1}}$ angstrom$^{\text{-2}}$ for 500 steps of SD minimization followed by 500 steps of CG minimization. Minimization Stage 2 is minimizing the entire system for 1000 steps of SD minimization followed by 1500 steps of CG minimization. The minimized models were checked by Swiss-PdbViewer and found there is not any bad contact that makes amino acids clashed. Then, the solvated proteins were quickly heated from 0 K to 310 K linearly for 300 ps and then systems were kept at 310 K for 700 ps. The systems were in constant NVT ensembles using Langevin thermostat algorithm with weak restraints (a force constant of 10.0 kcal mol$^{\text{-1}}$ angstrom$^{\text{-2}}$) on the solvated proteins. The SHAKE and PMEMD algorithms with nonbonded cutoffs of 12 angstroms were used. Next, the systems were done MD simulations in constant NPT ensembles (with constant pressure 1 atm and constant temperature 310 K) under the Langevin thermostat for 4 ns and the PRESS, VOLUME (DENSITY) and RMSDs were sufficiently stable for each of the 18 models (Figure 7). A step size of 2 fs was used for all the MD simulations, the structures were saved to file every 1000 steps and the Metropolis criterion was used. These MD simulation conditions are completely consistent with the experimental work of NMR structure of HuPrP(110--136) (2LBG.pdb).

\begin{figure}[h!] \label{Fig07}
\centerline{
\includegraphics[width=6.9in]{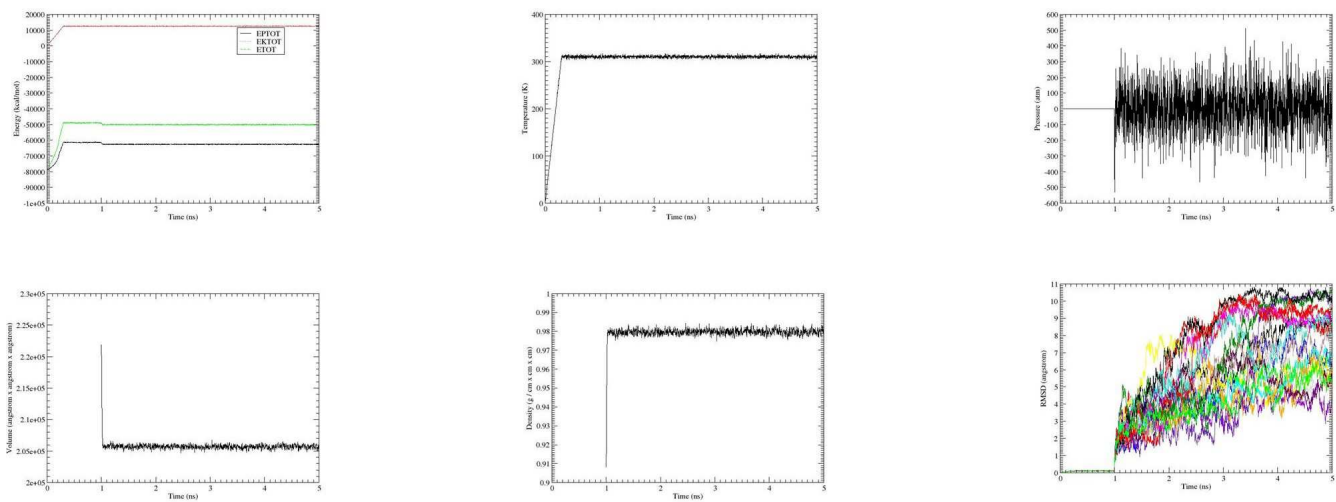}
}
\caption{{\it Variations of the potential energy (EPTOT), kinetic energy (EKTOT), total energy (ETOT), temperature, pressure, volume, density, and RMSDs during the 5 ns of MD simulations for all the models. In the RMSD graph, different colors stand for HuPrP(110--136) and its 17 mutants.}}
\end{figure}

We picked out the snapshots at 0, 1, 2, 3, 4 and 5 ns for each model (Figure 8) and found the $\alpha$-helix structure of each model has been unfolding and longer MD simulations might make the $\alpha$-helix structure unfolded completely (Figures 8$\sim$9). This shows that the structure in PrP(110--136) region is structurally unstable and might be critical to the conversion from the predominantly $\alpha$-helical PrP$^{\text{C}}$ into the rich $\beta$-sheet PrP$^{\text{Sc}}$.
\begin{figure}[h!] \label{Fig08}
\centerline{
\includegraphics[width=6.9in]{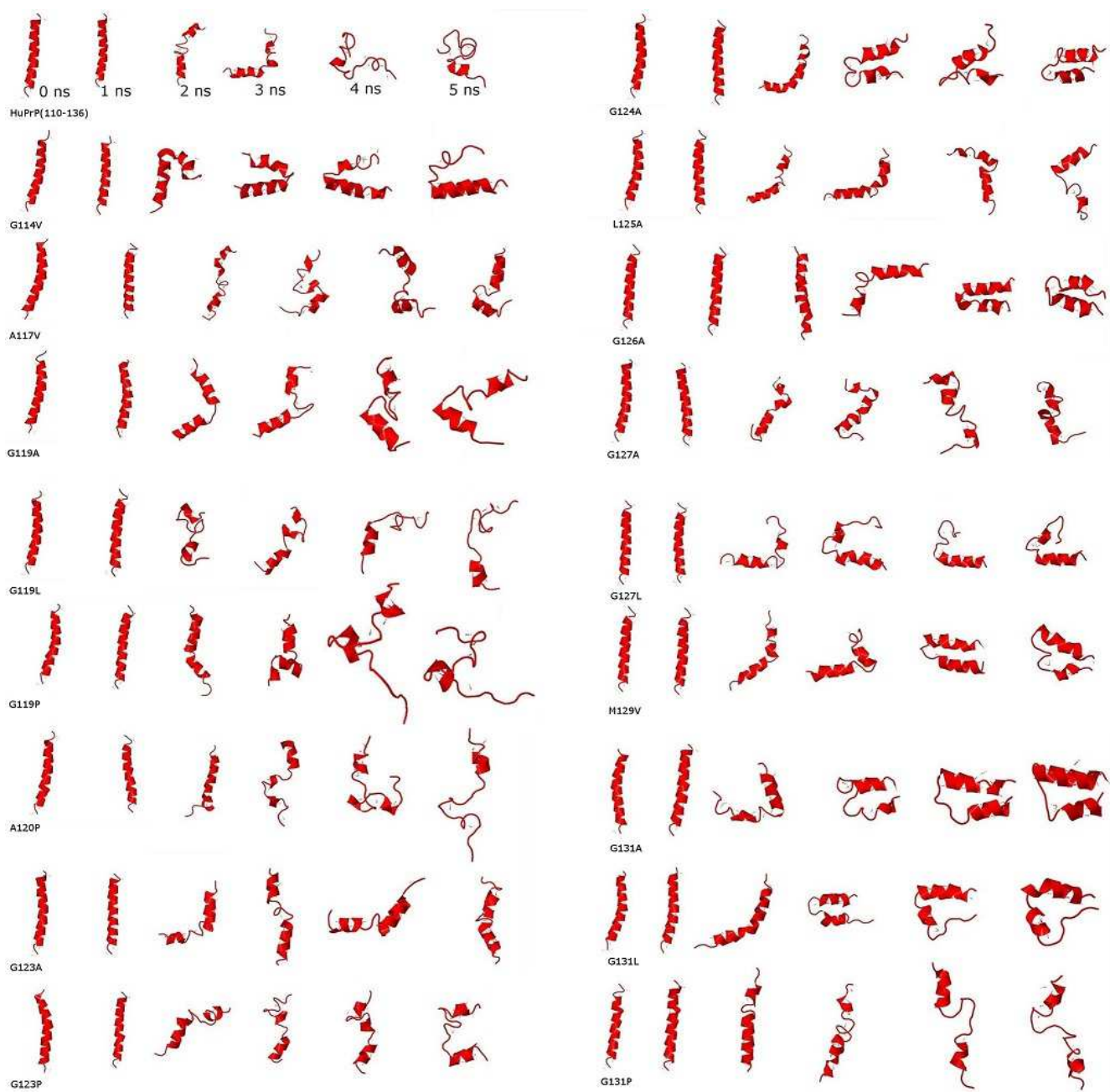}
}
\caption{{\it The respective snapshots at 0 ns, 1 ns, 2 ns, 3 ns, 4 ns and 5 ns of the MD simulations for the 18 models. The dashed lines denote hydrogen bonds.}}
\end{figure}
\begin{figure}[h!] \label{Fig09}
\centerline{
\includegraphics[width=6.9in]{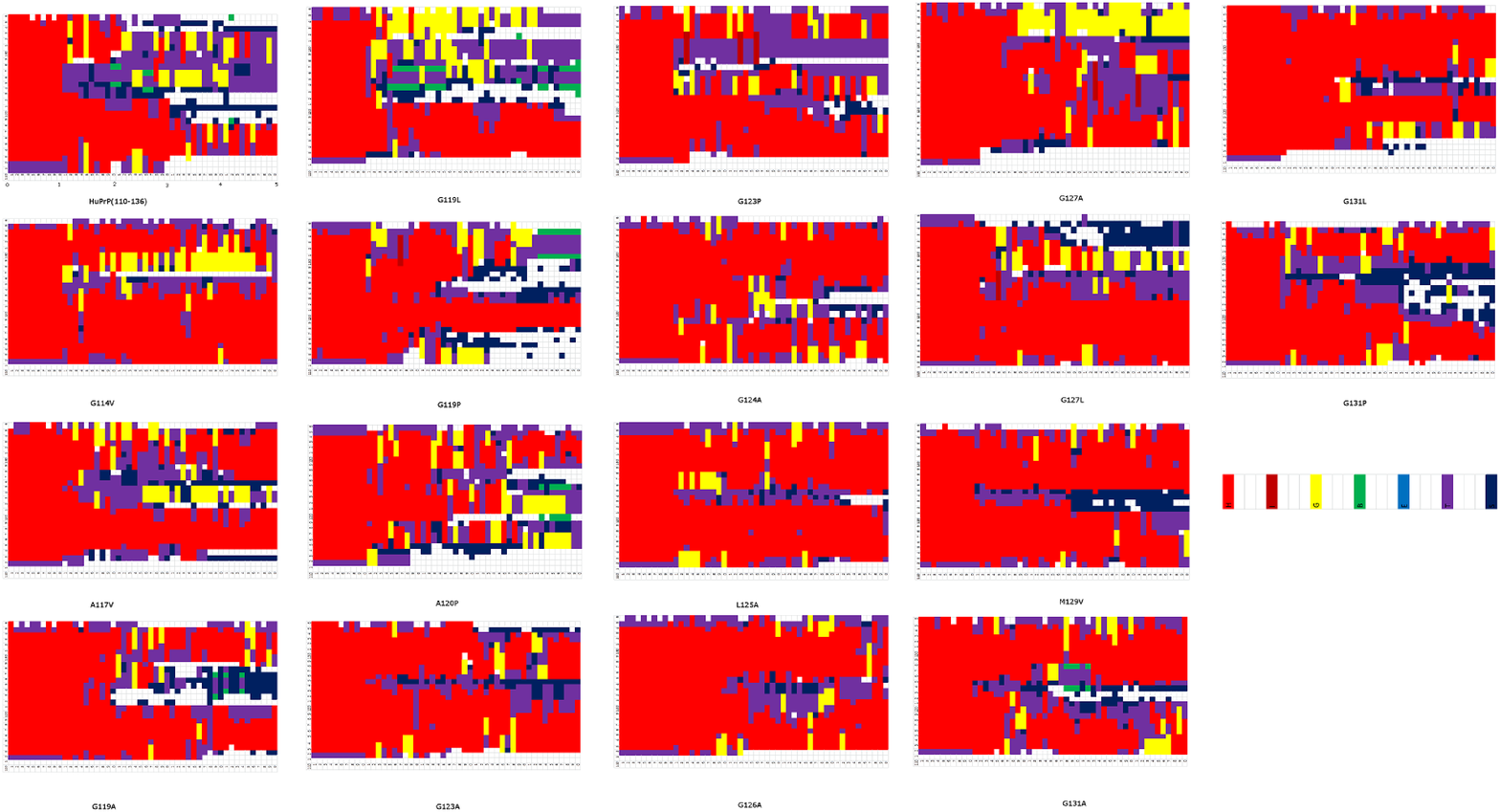}
}
\caption{{\it Variations of the secondary structures of the 18 models. H is the $\alpha$-helix, B is the residue in isolated $\beta$-bridge, E is the extended strand, G is the 3-helix or 3/10 helix, I is the $\pi$-helix, T is the hydrogen bonded turn and S is the bend as described in the DSSP program \cite{kabsch_etal1983}.}}
\end{figure}

The preliminary findings to explain the above performance during the 5 ns of MD simulations may be described as follows. At the beginning of the simulations, the 3D structures of all the 13 models are $\alpha$-helices (Figures 8$\sim$9). with rich HBs. In an ideal $\alpha$-helix, there are 3.6 residues per complete rotating so a rotation of 100 degrees per residue. Thus, in all the 13 models, there are about 7 turns (Figure 8) in each $\alpha$-helix. A $\alpha$-helix is maintained by HBs. Clearly, the disappearance of HBs is a reason for the unfolding of all the 13 models.  As the progress of MD simulations, most of them have disappeared, except for the three main HBs GLY/ALA/LEU/PRO131--SER135--ARG136, TYR128--SER132, which have different occupied rate for each model (Table 1) and are just in the N-terminal of the structural region of HuPrP(125--228) (1QLX.pdb). This might show that PrP(110--124) region of PrP(110--136) (2LBG.pdb) is very unstable.
\begin{table}[h]
\caption{Occupied rate (\%) of the hydrogen bonds for each model}
\centering
{\tiny
\begin{tabular}{c                            c                            c                          c} \hline \hline 
                HuPrP(110--136)              &G114V                       &A117V                     &G119A\\ \hline
                {\bf 131@O--135@OG.HG} (25.48)     &{\bf 131@O--135@OG.HG} (26.28)    &{\bf 131@O--135@OG.HG} (40.52)  &{\bf 131@O--135@OG.HG} (49.24)\\ 
                125@O--135@OG.HG (17.68)     &{\bf 135@O--136@NH1.HH11} (21.72) &{\bf 128@O--132@OG.HG} (8.88)   &{\bf 135@O--136@NH1.HH11} (14.6)\\
                117@O--136@NH1.HH11 (7.32)   &129@O--132@OG.HG (7.76)     &{\bf 135@O--136@NH1.HH11} (8.12)&{\bf 128@O--132@OG.HG} (5.68)\\
                116@O--136@NH1.HH12 (5.76)   &132@O--135@OG.HG (7.72)     &132@O--135@OG.HG (7.36)   &\\
\hline
\end{tabular}
\begin{tabular}{c                             c                           c                               c}  
                G119L                         &G119P                      &A120P                          &G123A\\ \hline
                {\bf 135@O--136@NH1.HH11} (23.96)   &{\bf 131@O--135@OG.HG} (31.24)   &135@O--136@OG.HG (28.08)       &{\bf 131@O--135@OG.HG} (28.08)\\ 
                {\bf 131@O--135@OG.HG} (22.8)       &{\bf 135@O--136@NH1.HH11} (24.4) &132@O--135@OG.HG (8.6)         &{\bf 135@O--136@NH1.HH11} (17.76)\\
                132@O--135@OG.HG (15.08)      &{\bf 128@O--132@OG.HG} (10.08)   &{\bf 135@O--136@NH1.HH11} (7.72)     &{\bf 128@O--132@OG.HG} (6.16)\\
                                              &130@O--135@OG.HG (7.2)     &112@O--111@ND1.HD1 (5.6)       &\\
                                              &                           &{\bf 128@O--132@OG.HG} (5.04)        &\\
\hline
\end{tabular}
\begin{tabular}{c                             c                              c                               c}  
                G123P                         &G124A                         &L125A                          &G126A\\ \hline
                {\bf 131@O--135@OG.HG} (25.52)      &{\bf 131@O--135@OG.HG} (41.4)       &{\bf 131@O--135@OG.HG} (36.24)       &{\bf 131@O--135@OG.HG} (46.36)\\ 
                {\bf 135@O--136@NH1.HH11} (21.92)   &{\bf 135@O--136@NH1.HH11} (21.4)    &{\bf 135@O--136@NH1.HH11} (17.92)    &{\bf 135@O--136@NH1.HH11} (14.12)\\
                                              &{\bf 128@O--132@OG.HG} (5.56)       &132@O--135@OG.HG (8.48)        &{\bf 128@O--132@OG.HG} (7.04)\\
                                              &134@O--136@NH1.HH11 (5.44)    &{\bf 128@O--132@OG.HG} (5.6)         &133@O--136@NE.HE (5.6)\\
                                              &                              &                               &132@O--135@OG.HG (5.08)\\
\hline
\end{tabular}
\begin{tabular}{c                             c                              c                               c}  
                G127A                         &G127L                         &M129V                          &G131A\\ \hline
                {\bf 131@O--135@OG.HG} (26.2)       &{\bf 135@O--136@NH1.HH11} (23.6)    &{\bf 131@O--135@OG.HG} (34.4)        &{\bf 131@O--135@OG.HG} (47.76)\\ 
                {\bf 135@O--136@NH1.HH11} (14.88)   &{\bf 131@O--135@OG.HG} (21.2)       &{\bf 135@O--136@NH1.HH11} (13.2)     &{\bf 128@O--132@OG.HG} (17.08)\\
                112@O--111@ND1.HD1 (13.08)    &133@O--135@OG.HG (8.04)       &{\bf 128@O--132@OG.HG} (8.92)        &{\bf 135@O--136@NH1.HH11} (13.48)\\
                                              &                              &                               &132@O--136@NH1.HH11 (7.72)\\
                                              &                              &                               &134@O--111HIS@ND1.HD1 (6.2)\\
\hline
\end{tabular}
\begin{tabular}{c                             c}  
                G131L                         &G131P\\ \hline
                {\bf 131@O--135@OG.HG} (70.4)       &{\bf 131@O--135@OG.HG} (32.48)\\
                {\bf 135@O--136@NH1.HH11} (18.4)    &{\bf 135@O--136@NH1.HH11} (16.72)\\
                {\bf 128@O--132@OG.HG} (7.88)       &{\bf 128@O--132@OG.HG} (15.2)\\
\hline
\end{tabular}
}
\end{table}
\noindent We also found that there is a SB between HIS111--LYS110 with the occupied rate 100\% for the 13 models. Specially for the mutants G127L, M129V, G131A, and G131L, there is another SB between HIS111--136ARG linking the head and tail of HuPrP(110--136). Seeing the snapshots of 3, 4 and 5 ns of mutant G127L in Figure 8, 4 and 5 ns of mutant M129V in Figure 8, 3, 4 and 5 ns of mutant G131A in Figure 8, and 3, 4 and 5 ns of mutant G131L in Figure 8, we may know the SB HIS111--136ARG makes these snapshots looking like  a ``hairpin". HIS111 is a very important residue in PrP(110--136). Along with the unfolding of $\alpha$-helical structure of all these 13 models, we found many HPs disappeared except for some fundamental ones MET134--ALA133, LEU130--129MET, VAL122--VAL121-ALA120, ALA118--ALA117--ALA116--ALA115, and ALA113--MET112 with the occupied rate 100\%, where ALA118--ALA117--ALA116--ALA115 are in the core of the palindrome AGAAAAGA and this might imply to us the hydrophobic core is very hard to break and this palindrome really has enormous potential to be amyloid fibrils. 

\section{Concluding Remarks on PrP(109--136)}
To really reveal the secrets of prion diseases is very hard. For us it is a long shot but certainly worth pursuing. It was reported that the hydrophobic region PrP(109--136) controls the formation into diseased prions: the AGAAAAGA palindrome and Glycine-xxx-Glycine repeats (both being the inhibitor of prion diseases) are just in this region. This paper gives some investigations and explanations on the PrP(109--136) region in view of its 3D structures and molecular dynamics studies. The structural bioinformatics presented in this paper can be acted as a reference in 3D images for laboratory experimental works. This presents some clue or hints for the author to study prion proteins and prions. 

\section*{Acknowledgments}
{\small This research has been supported by a Victorian Life Sciences Computation Initiative (VLSCI) grant numbered VR0063 on its Peak Computing Facility at the University of Melbourne, an initiative of the Victorian Government of Australia.}

\end{document}